\documentclass[onecolumn,11pt,showpacs,superscriptaddress,preprintnumbers,amsmath,amssymb,aps,prl,floatfix]{revtex4-1}
\pdfoutput=1
\usepackage{graphicx}
\usepackage{bm}
\usepackage{color}
\usepackage{verbatim}
\usepackage{textcomp}
\newcommand{\beq}{\begin{equation}}
\newcommand{\eeq}{\end{equation}}
\newcommand{\bea}{\begin{eqnarray}}
\newcommand{\eea}{\end{eqnarray}}

\newcommand{\heidelberg}{Physikalisches Institut, Universit{\"a}t Heidelberg, Philosophenweg~12, 69120 Heidelberg, Germany}
\newcommand{\tum}{Technische Universit{\"a}t M{\"u}nchen, Franck-Hertz-Stra{\ss}e, 85748 Garching, Germany}
\newcommand{\frm}{FRM II, Technische Universit{\"a}t M{\"u}nchen, Lichtenbergstra{\ss}e 1, 85748 Garching, Germany}
\newcommand{\ill}{Institut Laue-Langevin, BP~156, 6,~rue Jules Horowitz, 38042 Grenoble Cedex~9, France}
\newcommand{\ati}{Atominstitut, Technische Universit{\"a}t Wien, Stadionallee~2, 1020 Wien, Austria}

\usepackage[	pdftex,
			colorlinks=false,
			final,
			unicode,
			pdftitle={A quantized frequency reference in the short-ranged gravity potential and its application for dark matter and dark energy searches},
			pdfauthor={T. Jenke, G. Cronenberg, P. Geltenbort, A.N. Ivanov, T. Lauer, T. Lins,  H. Saul, U. Schmidt, H. Abele}]
		{hyperref}
	
\begin{document}

\title{A quantized frequency reference in the short-ranged gravity potential and its application for dark matter and dark energy searches}

\author{T. Jenke}\email{tjenke@ati.ac.at} \affiliation{\ati}
\author{G. Cronenberg} \affiliation{\ati}
\author{P. Geltenbort} \affiliation{\ill}
\author{A.N. Ivanov} \affiliation{\ati}
\author{T. Lauer} \affiliation{\frm}
\author{T. Lins} \affiliation{\tum}
\author{U. Schmidt} \affiliation{\heidelberg}
\author{H. Saul} \affiliation{\ati}
\author{H. Abele}\email{abele@ati.ac.at} \affiliation{\ati}

\date{\today}


\begin{abstract}
The LHC experiments have reached a milestone in our understanding of nature. The evidence for the observation of the Higgs spin-0-boson as a manifestation of a scalar field provides the so far missing corner stone for the standard model of particles and fields, "which describes the fundamental particles from which every visible part of the universe are made of and the forces between them"~\cite{LHC2012,atlas,cms}. However, the standard model fails to explain the other, non-visible but gravitationally active part of the universe, contributing to about $95\%$ of the dark energy/mass density of the universe. Its nature is unknown but the confirmation of a scalar Higgs is giving a boost to scalar-field-theories in an attempt to explain the other, non-visible part of the universe. So far gravity experiments and observations performed at different distances find no deviation from Newton's gravity law and also fail to find a fifth force contributed to dark energy or dark matter~\cite{Adelberger2009}. Therefore dark energy must possess a screening mechanism which suppresses the scalar-mediated fifth force. Our line of attack is a novel gravity experiment with neutrons based on a quantum interference technique. In particular the spectroscopic measurement of quantum states on resonances with an external coupling makes this search in the gravity potential of the Earth a powerful way to search for dark matter and dark energy contributions in the universe. Quantum states in the gravity potential are intimately related to other scalar field or spin-0-bosons if they exist.  If the reason is that some undiscovered particle interact with a neutron, this should result in a measurable energy shift of quantum states in the gravity potential of the Earth, because for neutrons the screening effect is absent~\cite{Brax2011}. Here we use Gravity Resonance Spectroscopy~\cite{Jenke2011} to measure the energy splitting at the highest level of precision reported so far, providing a constraint on any possible new interaction. We obtain a sensitivity of 10$^{-14}$ eV. A fit to the data sets an upper limit on any fifth force, in particular on parameter $\beta < 2\times 10^9$ at $n = 3$ for the scalar chameleon field, which is improved by a factor of 100 compared to the upper bound~\cite{Brax2011} derived from our previous experiment~\cite{Nesvizhevsky2005,Westphal2007} and five orders of magnitude better than the upper bound, $\beta < 10^{14}$, from precision tests of atomic spectra~\cite{Brax2011}. The pseudoscalar axion coupling is constrained to $g_sg_p/\hbar c < 3 \times 10^{-16}$ at $20~${\textmu}m, which is an improvement by a factor of 30. These results, consistent with zero, indicate that gravity is understood at this improved level of precision.
\end{abstract}
\maketitle

The last fourteen years of experiments and observation~\cite{Riess1998} have produced an entirely new map of the universe. Cosmological parameters describe the properties and the global dynamics of the universe. Observational cosmology has determined these parameters to one or two significant figure accuracy. Of great interest is now how the energy-matter budget is built up from its constituents: e.g. baryons, photons, neutrinos, dark matter and dark energy~\cite{Lahav2010}. The known particles of the Standard Model~\cite{Glashow1970,Weinberg1967,Salam1969} account for only 5$\%$, whereas the majority consists of unknown dark energy and dark matter.

Firstly, dark energy -- interpreted as negative pressure -- is causing the universe to accelerate. A particularly attractive possibility is that a scalar field, usually called quintessence, is responsible for inflating the universe~\cite{Silvestri2009,Caldwell2009,Copeland2006,Wetterich1988}. A competing possibility is that the acceleration is due to a modification of gravity, i.e., the left-hand side of Einstein's equation rather than the right. Observations, which can distinguish these two possibilities are desirable, since measurements of expansion kinematics alone are not able to. Remarkably, the best limits for scalar fields (chameleons) come from our experiment with ultra-cold neutrons, as we will show.

Secondly, distinct from dark energy is the concept of dark matter observed indirectly, through the influence of its gravity on surrounding ordinary matter. It has evaded more conventional observation using optical, radio or other telescopes, so it evidently only has a feeble coupling to photons. We do not yet know precisely what this dark matter is, but the leading theoretical candidates (axions and weakly interacting massive particles, WIMPs) are massive particles that do not acquire their mass by interacting with the electroweak Higgs condensate. One eye-catching concept is the idea of detecting very light bosons through the macroscopic forces which they mediate. These particles may be responsible for the dark matter in the universe or part of it. They must be very light with a mass of about 10~{\textmu}eV to 1~eV to be compatible with astronomical observation \cite{Beringer2012}. This corresponds to a range of 0.2~{\textmu}m and 2~cm they can mediate, leading to a deviation from Newton's law at short distances, exactly in the range of this experiment.

Thirdly, there are indications that all known basic forces, the weak force, the strong force, electromagnetism and gravity, might be unified at high energies. One approach are string-theories, which are naturally defined in ten or eleven space-time dimensions. The hope is that a future theory of quantum gravity can be constructed in such a way that all forces, including gravity, could be combined within it. Such a theory cannot be constructed in a consistent way if space-time is limited to four dimensions, so additional spatial dimensions are needed to accommodate such a theory in a reasonable way. Extra-dimensions have not been observed in everyday life, because these extra-dimensions might be "curled-up" with a possibly very small radius of curvature. This should lead to deviations from Newton's gravitational law at very small distances. The focus of these has shifted over the years and it has been realized, that the string scale, which controls the strength of gravity as well as the strength of other forces, is not necessarily related to the Planck length of 10$^{-35}$~m, instead it can be considered as a free parameter and we might live with large extra dimensions~\cite{Arkani-Hamed1999}, again in the range of our experiments~\cite{Abele2003}.

The idea of finding a common ground to the problems related to string theories, dark energy and dark matter is aesthetically appealing to us. That these problems might have a common solution can be seen that these approaches are strongly interrelated, not only by a unified physics aim, like the role of gravity in modern gauge theory for cosmology, but also by the common methodical approach addressed by this experiment.

All these ideas triggered gravity experiments of different kinds, which in the past ten years have validated Newton's gravitational law down to about 20~{\textmu}m~\cite{Fischbach1999,Adelberger2009,Tu2007,Abele2003,Hoyle2004,Decca2005,Long2003,Geraci2008,Chiaverini2003,Bordag2001}. Searches for short-ranged non-Newtonian gravity so far have been performed using neutrons, atoms, torsion pendulums, and other macroscopic objects.
To date, neutron scattering experiments provide most stringent limits at very short distances around {1-10}~nm~\cite{Leeb1992,Nesvizhevsky2008}. A recent review on "the neutron and its role in cosmology and particle physics" can be found in Ref.~\cite{Dubbers2011a}. Currently, measurements with atomic force microscopes determine the limits on non-Newtonian gravity below 10~{\textmu}m.
\begin{figure}[t]
	\centering
	\parbox{0.45\textwidth}{
		\includegraphics[width=0.45\textwidth]{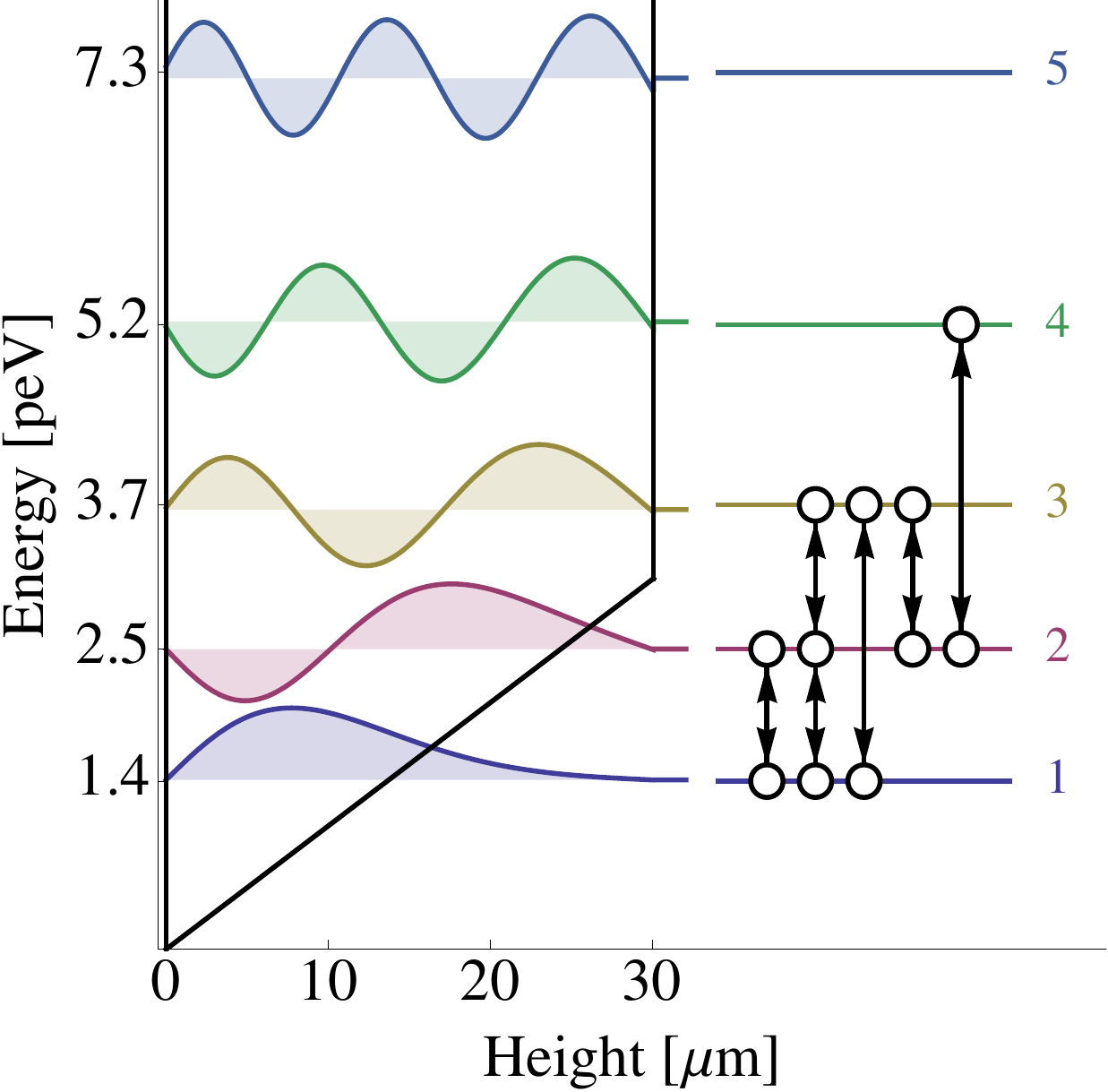}
	}
	\parbox{0.45\textwidth}{
		\begin{minipage}{0.45\textwidth}
			\vspace{1.6cm}
			\begin{tabular}{|l|ccccc|}
			\hline
			&2&3&4&5&6\\\hline
			1	&	262,7 & 542,1 & 918,5 & 1405,3  & 2003,4  \\
			2	&          & 279,4 & 655,4 & 1142,6  & 1740,7  \\
			3	&       &          & 376,0 & 863,2   & 1461,3  \\
			4	&       &       &          & 487,3   & 1085,3  \\
			5	&       &       &       &             & 598,1   \\
			\hline
			\end{tabular}
		\end{minipage}
	}
	\caption{Left: Frequency reference for gravity experiments based on energy eigenvalues of a neutron in the gravity potential of the Earth with confining mirrors at the bottom and at the top separated by 30.1~{\textmu}m. Also shown is the neutron density distributions for energy levels $|1\rangle$ to $|5\rangle$. The Gravity Resonance Spectroscopy technique allows to drive transitions between these states. Quantum transitions $|1\rangle \leftrightarrow|2\rangle$, $|1\rangle \leftrightarrow|3\rangle$,  $|2\rangle \leftrightarrow|3\rangle$, and $|2\rangle \leftrightarrow|4\rangle$ have been observed.\newline
Right: Resonant transition frequencies in Hz at width $l = 30.1$~{\textmu}m between mirrors, see Methods.}
	\label{fig:En}
\end{figure}
\begin{figure}[t]
	\centering
		\includegraphics[width=0.45\textwidth]{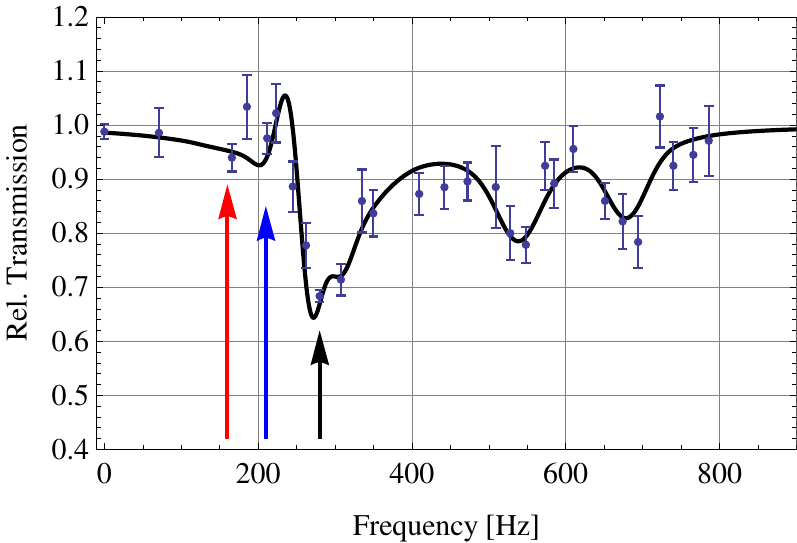}
		\includegraphics[width=0.45\textwidth]{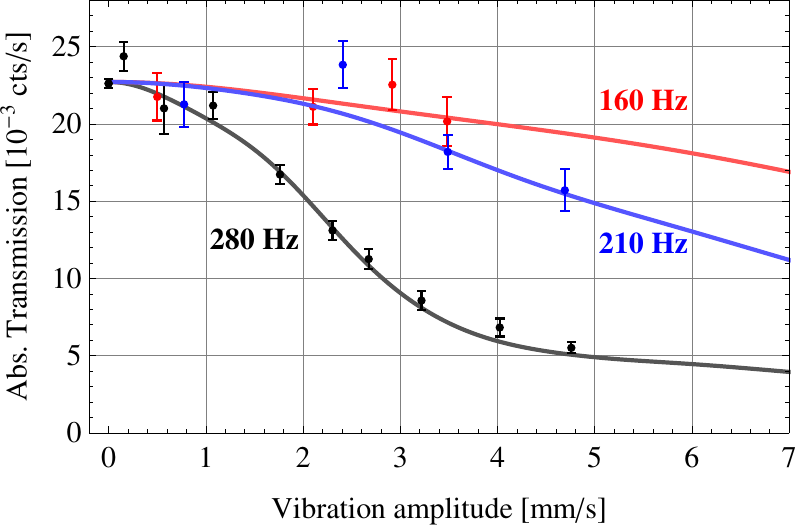}
	\caption{Gravity Resonance Spectroscopy:\newline
Left: Transmission as a function of frequency. Transitions between $|1\rangle \leftrightarrow|2\rangle$ , $|1\rangle \leftrightarrow|3\rangle$, $|2\rangle \leftrightarrow|3\rangle$, and $|2\rangle \leftrightarrow|4\rangle$ have been observed. We find transmission frequencies $\nu_{12} = \left(258\pm 2\right)$~Hz, $\nu_{23} = 280^{+3}_{-2}$~Hz, $\nu_{13} =  \left( 539\pm3 \right)$~Hz, $\nu_{24} =  679^{+13}_{-15}$~Hz.\newline
Right: Damped Rabi-oscillations and Transmission as a function of amplitude strength in resonance and at two other positions in the rising slope, see arrows to the left. The resonant frequencies have been determined to up to $5 \times 10^{-3}$ corresponding to an energy sensitivity of  $E = 10^{-14}$~eV.}
	\label{fig:GRS}
\end{figure}

Such experiments clearly show the basic problem in searching for new physics at small distances: that the size of the objects under study needs to be reduced, which is accompanied by a reduction in signal intensity. At the same time, the electrostatic background increases.  The gravitational interactions are studied in the presence of large van-der-Waals and Casimir forces, which depend strongly on the geometry of the experiment, and the theoretical treatment of the Casimir and other polarizability effects is a difficult task. In practice, the experimental data are subject to corrections, which can be orders of magnitude larger than the effects actually searched for~\cite{Adelberger2009}.

Our solution to this problem is the development of a purely quantum mechanical system for gravity measurements, where polarizability effects for all practical reasons are absent, i.e. the excitation of bound quantum states of a neutron in the gravity potential of the Earth, and its interaction with a macroscopic system, here a mirror for neutron reflection. This so-called Gravity Resonance Spectroscopy technique~(GRS) does not rely on electromagnetic fields or a coupling to an electromagnetic potential~\cite{Jenke2011,Jenke2011a}.

To improve on this resonance techniques we have developed a novel frequency reference system for gravity measurements, see Fig.~\ref{fig:En}. This system has two great advantages: Firstly, the reference system is only based on natural constants: the mass of the neutron~$m$, the Planck constant~$\hbar$, and the acceleration of the Earth~$g$. The constants~$m$ and $\hbar$ are known within 10$^{-8}$ and $g$ can be measured with a corner cube within $\Delta g/g$ = 2 $\times$ 10$^{-9}$. Tidal changes affect $g$ within 10$^{-7}$. Such a notional level of accuracy for the energy spacing in the gravity potential of the Earth is a challenge for our experiment to be probed by GRS. It corresponds to an energy change of around  $\Delta E$ = 10$^{-20}$~eV. On this level of accuracy, we want to apply GRS to the measurements of the inverse-square law of the Newton gravity at length scales of the order of a micron, where it is a solid testing-ground for theoretical schemes of interactions, based on supersymmetric theories of strings and D-branes, predicting possible deviations from Newton's Inverse Square Law. The lowest energy eigenvalues $E_n$, ($n \geq 1$) are on the pico-eV energy-scale. In this experiment these energy eigenvalues are tuned by a second mirror on top of the bottom mirror at a height of 30.1~{\textmu}m as described in Ref.~\cite{Jenke2011}. This changes the resonance frequency between $|p\rangle$ and $|q\rangle$ due to the additional mirror potential, which shifts the energy of states $n >  1$, but leaves state $|1\rangle$ unchanged. The additional potential provides an effective increase in sensitivity by a factor of two in a search for hypothetical axion or chameleon induced phase shifts or for other fifth forces. Secondly, we have now on hand precision measurements of resonant transitions between states $|1\rangle \leftrightarrow|2\rangle, |1\rangle \leftrightarrow|3\rangle, |2\rangle \leftrightarrow|3\rangle, $ and $ |2\rangle \leftrightarrow|4\rangle$ in the frequency regime between 70~Hz and 800~Hz. This quantum system represents a self-consistent frequency system, where systematic effects are virtually absent. It can be compared -- however with much less precision so far -- with a frequency reference in laser or microwave spectroscopy, e.g. atomic clocks, nuclear magnetic resonance techniques or M{\"o}\ss bauer spectroscopy, where the observables have generally been restricted to electromagnetic interactions. Our current sensitivity is  $E = 10^{-14}$~eV, see Table~\ref{tab:freqfiterg} and Fig.~\ref{fig:GRS}.

Fig.~\ref{fig:lins} shows the apparatus consisting of a mirror underneath and a mirror on top at distance~$l$ with rough surface. The neutrons are taken from the ultra-cold neutron installation PF2 at Institut Laue-Langevin (ILL). The experiment itself is mounted on a polished plane granite table with a passive anti-vibration tables underneath. The granite is leveled to an accuracy of better than $10$~{\textmu}rad and stabilized with a precision better than 1~{\textmu}rad. A mu-metal shield suppresses the coupling of residual fluctuations of the magnetic field to the magnetic moment of the neutron sufficiently. Gravity Resonance Spectroscopy is a Rabi-type spectroscopy method and consists of a state selector, where the initial state $|p\rangle$ is prepared. Transitions into a second state $|q\rangle$ are induced by applying  a so-called  $\pi$-pulse. A detector at the end measures these neutrons, which were accepted by the state selector. On resonance a sharp decrease in count rate is found. The frequency width of the resonance is given by the time the neutron spends in the oscillating system.

In general, the oscillator frequency at resonance for a transition between states with energies $E_p$ and $E_q$ is given by
\beq
	\nu_{pq}=\frac{1}{2\pi}\frac{E_q-E_p}{\hbar}.
\label{fig:omega}
\eeq
The linear gravity potential leads to discrete energy eigenstates of a bouncing quantum particle above a horizontal mirror. Neutrons are trapped by gravity, but the motion is unrestricted parallel to the mirror.  So far the systematic effects of the top mirror on the resonant transitions are smaller than our statistical error, see Methods.
A measured energy level splitting can be compared with the prediction in the gravity potential, allowing a test of Newton's law at short distances. The lowest energy eigenvalues $E_n$ read 1.41~peV, 2.50~peV, 3.66~peV, 5.22~peV, and 7.25~peV. Another GRS measurement, based on oscillating magnetic gradient fields, is in progress~\cite{Kreuz2009}.

Our measurement is derived from 116 neutron transmission measurements taken in 2010 and 2011, together with many subsidiary measurements used to search for systematic errors. The neutron transmission in counts per second as a function of frequency and amplitude is registered. Included are 17~measurements, where the vibration amplitude is zero. After background subtraction of $\left(2.18 \pm 0.08\right)\times 10^{-3}$~s$^{-1}$, the following 10~parameters fit the 116~data points, using Eq.~\ref{eq:fitvib01}, see Tab.~\ref{tab:freqfiterg}: 
Firstly, three parameters $\nu_{12}$, $\nu_{23}$, and $\nu_{24}$ correspond to the transition frequencies. 
Energy conservation requires  $\nu_{12}$ + $\nu_{23}$ = $\nu_{13}$. 
Secondly, three parameters $\gamma_2$, $\gamma_3$, and $\gamma_4$ describe the damping of the  probability amplitude of the states due to the rough, upper surface of the neutron mirror with respect to the damping of the ground state $1\rangle$. 
Thirdly, the initial populations $c_2$, $c_3$ and $c_4$ for state $2\rangle$ to $4\rangle$ with respect to state $1\rangle$ and an overall normalization. 
In a first analysis, resonant transitions $\nu_{12}$, $\nu_{23}$, and $\nu_{24}$ are determined.  $\nu_{13}$ is obtained from energy conservation. A $\chi^2_{min}$ = 114.7 at 116 data points and 10~parameters is found, which corresponds to a probability of 26.6$\%$. Resonant frequencies are determined with a precision of $8 \times 10^{-3}$ for transition  $1\leftrightarrow2$,  1$\%$ for transition  $2\leftrightarrow3$, $5 \times 10^{-3}$ for transition  $1\leftrightarrow3$, and 2$\%$  for transition  $2\leftrightarrow4$, see Tab.~\ref{tab:freqfiterg}.
In fact, the resonant transitions are given by the acceleration of the Earth and the slit height~$l$ between the two mirrors. Due to the reduced number of parameters, the  $\chi^2$ is worse, but with $g = 9.81$~m/s$^2$, which is the value for Grenoble, the probability is still 16.7$\%$.

Fig.~\ref{fig:GRS}, left, shows the measured relative transmission. Clearly visible is a decrease of transmission as a function of frequency. Such dips correspond to relevant transition frequencies  $\nu_{12}$, $\nu_{13}$, $\nu_{23}$, and $\nu_{24}$. At a frequency $\nu = 280$~Hz, the absolute resonance minimum is found. Scanning the vibration amplitude generates a damped Rabi-oscillation curve as predicted, shown in Fig.~\ref{fig:GRS}, right. These measurements altogether can be displayed in one single data point corresponding to one specific frequency (black arrow in Fig.~\ref{fig:GRS}, left), by normalizing the measured count rate to the average rate without vibration multiplied by the best fit for a chosen vibration strength divided by the best fit at that point. The instruction offers an elegant way in displaying all 116 measurements in one Figure.~\ref{fig:GRS}, left, irrespective of vibration strength. For these relative transmission points a binning of 20 Hz is used. The chosen vibration strength is 2.03 mm/s, which is the average of all vibration measurements.

\begin{table}[tb]
	\centering
	\begin{tabular}{||l|r|r|r||}
	\hline\hline
Parameter & best fit & -error & +error \\
\hline
$\nu_{12} \mathrm{[Hz]}$ & 258 & -2 & 2 \\
$\nu_{23} \mathrm{[Hz]}$ & 280 & -2 & 3 \\
$\nu_{13} \mathrm{[Hz]}$ & 539 & -3 & 3 \\
$\nu_{24} \mathrm{[Hz]}$ & 679 & -16 & 13 \\
$\gamma_2$ & 85.5 & -5.6 & 6.3 \\
$\gamma_3$ & 312 & -13 & 15 \\
$\gamma_4$ & 255 & -72 & 185 \\
$|c_2|^2$  & 5.8 & -1.2 & 0.9 \\
$|c_3|^2$  & 0.1 & -- & 0.6 \\
$|c_4|^2$  & 0.0 & -- & 3.4\\
 \hline\hline
\end{tabular}
\caption{Best parameter set at $\chi^2_{min}$ and error at 1$\sigma$ standard deviation ($\chi^2_{min} +1)$ in 10 dimensional hyperspace. }
	\label{tab:freqfiterg}
\end{table}
\begin{figure}[!htb]
	\centering
		\includegraphics[width=0.45\textwidth]{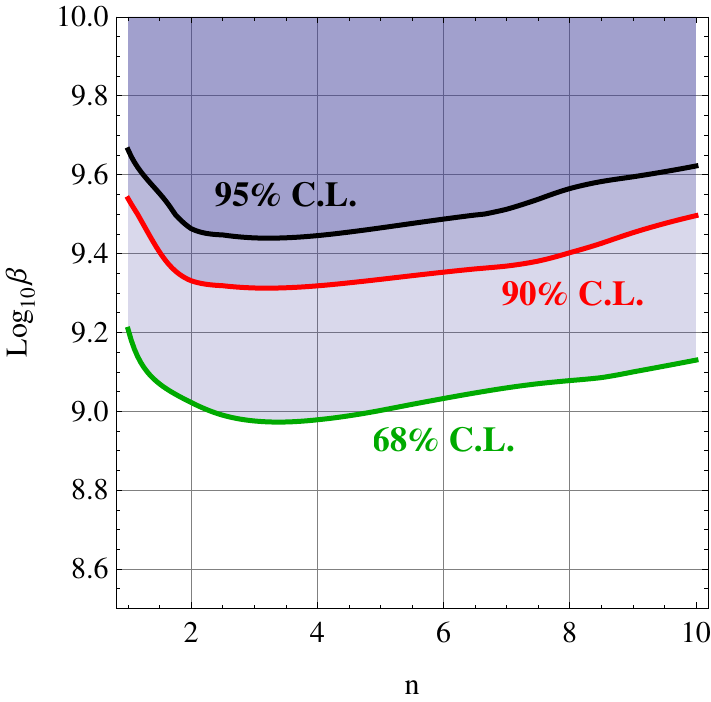}
		\includegraphics[width=0.45\textwidth]{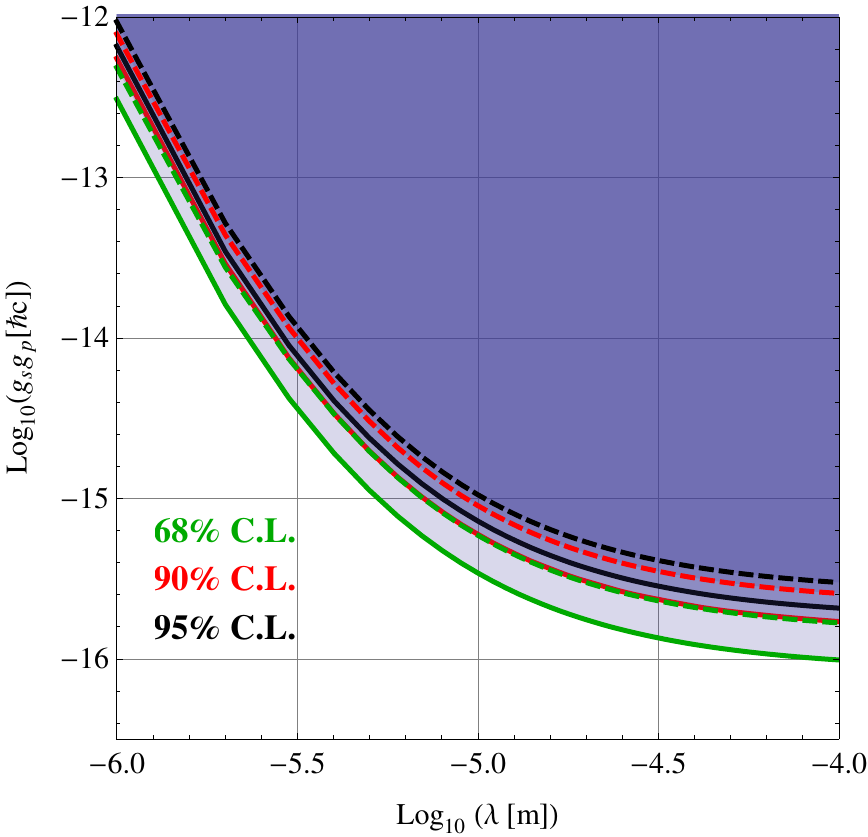}
	\caption{Exclusion plots with 68$\%$, 90$\%$, and 95$\%$ C.L.. Left, chameleon field: limits are shown as contours for parameter $\beta$ vs. parameter $n$, which is improved by a factor of 100~\cite{Brax2011} compared to the upper bound derived from our previous experiment~\cite{Nesvizhevsky2005,Westphal2007} and 5 orders of magnitude better than the upper bound, $\beta$ $<$10$^{14}$, from precision tests of atomic spectra~\cite{Brax2011}. Right, axion field: limits are shown as contours for parameter $g_sg_p/\hbar c$  vs. range $\lambda$. A limit for an attractive coupling is shown in a solid line, whereas a repulsive force limit is indicated by the dashed line.}
	\label{fig:chameleonexclusion}
\end{figure}
In this section we perform a search for so called chameleon fields. A chameleon field  has been suggested to drive the current phase of cosmic acceleration for a large class of scalar potentials. The properties depend on the density of a matter to which it is immersed. Because of this sensitivity on the environment, the field was called chameleon, and it can couple directly to baryons with gravitational strength on Earth but it would be essentially massless on solar system scales \cite{Khoury2004,Mota2006,Mota2007,Waterhouse2006}. The coupling constant is usually referred to as $\beta$. Such a chameleon coupling to our neutron would provide a frequency shift of resonant transitions, which is proportional to $\beta$  and can be measured by comparing the transition frequency  $\nu_{pq}$ with its theoretical expectation  $\nu_{pq}^{theo}$, when no chameleon field is present. Our method directly tests the chameleon matter interaction and does not rely on the existence of a chameleon-photon-interaction as other
experiments do~\cite{Ahlers2008,Gies2008,Steffen2010,Upadhye2012,Rybka2010,Wester2011}.

The calculation of the contribution of a chameleon field to the transition frequencies of the quantum gravitational states of ultra-cold neutrons is related to the calculation of the matrix elements $\langle p|\phi(z)\rangle$, where $|p\rangle$  is a low--lying quantum state of ultra-cold neutrons in the gravitational field of the Earth. Due to the Airy functions describing the gravitational states of ultra-cold neutrons, the main contribution to matrix elements $\langle p| z |p \rangle$ comes from the region around 
$(d/2 + z)~l_0 = (2m^2g/\hbar^2)^{-1/3}$ = 5.9~{\textmu}m, which is a natural scale for neutron quantum states. In our case with two mirrors at distance $d$, the chameleon field potential is given by \cite{Ivanov2012}:
\begin{align}
\phi_{\rm ges}(z) &= mg\left(z + \frac{d}{2} \right) + \beta \frac{m}{M_{Pl}} \phi_0 \left(1 - \frac{4 z^2}{d^2} \right)^{\frac{2}{n+2}} \\
\phi_0 &= \Lambda \left( \frac{n+2}{4\sqrt{2}} \Lambda d\right)^{\frac{2}{n+2}},
\end{align}where $M_{Pl}$ is the Planck mass, the scale $\Lambda$   is chosen to be equal $\Lambda$   = 2.4(1) $\times$ 10$^{-12}$ GeV.  Quanta of a chameleon field are massive with mass defined by the second derivative of the potential $V(\phi)$. The coordinate system has been shifted by $d/2$ due to symmetry reasons. In a fit to the data, the Earth's acceleration has been fixed at $g = 9.81$~m/s$^{2}$, the other parameters have been varied. The extracted confidence intervals for limits on the parameters $\beta$  and $n$ can be found in Fig.~\ref{fig:chameleonexclusion}. The experiment is most sensitive at n=3 and excludes a coupling to chameleon fields with  $\beta > 2.1 \times 10^9$~(90$\%$~C.L.). This is a factor of 100 more precise than the previous upper limit~\cite{Brax2011}.

Second, we perform a search for spin-dependent forces at short distances. Resonant transitions are for this purpose measured with polarized neutrons. The experiment is therefore slightly modified: a homogeneous magnetic guiding field of 50~{\textmu}T preserves the neutron spin throughout the experiment. Neutron spin polarization is analysed by a soft ferromagnetic foil, which was coated on the detector foil for that purpose. A strong magnetic field at the position of the detectors is oriented parallel or antiparallel to the direction of gravity. By altering the direction of this field the analysing power with respect to the spin state direction and gravity direction can be changed using the hysteresis of the iron. A hypothetical spin-dependent force would change the transition frequencies. Such a frequency shift is measured by reversing the direction of the applied guiding field and the detector field and measuring the difference in the two signals. As a check, the signal is measured at two transition frequencies. This difference is positive on one side of the transmission curve and negative on the other. The most sensitive places are those with large slopes. We do not observe a significant frequency shift. Limits for spin-dependent forces can be easily interpreted as bounds of the strength of the matter coupling of axions. Axion interactions with a range within 0.2~{\textmu}m~$< m_a <$~2~cm (corresponding to axion masses $\left( 10^{-5} < m_a < 1\right)$~eV), the "axion window", are still allowed by the otherwise stringent constraints posed by cosmological data. An axion would feel a CP-violating spin-dependent interaction in the presence of matter given by~\cite{Moody1984}.
\beq V(\vec{r})=\hbar g_p g_s
\frac{\vec{\sigma}\cdot\vec{n}}{8\pi m c}\left(\frac{1}{\lambda
r}+\frac{1}{r^2}\right)\,e^{-r/\lambda}\;\;.\label{axpot1}\eeq
Here, $\vec{\sigma}$ denotes the neutron spin and $\vec{n}$ is a
unit vector related to the geometry of the macroscopic
matter configuration.

A fit of strength $g_sg_p/\hbar c$ and range $\lambda$  together with the other independent parameters, this leads at 90\%~C.L. to an upper limit on the axion interaction strength as shown in Fig.~\ref{fig:chameleonexclusion}, right. For example, at $\lambda = 20$~{\textmu}m in the astrophysical axion window, an attractive coupling strength $g_sg_p/\hbar c < 3 \times 10^{-16}$ is excluded. This is the most precise upper limit from a direct search for attractive and repulsive $g_sg_p/\hbar c$-coupling at this length scale $\lambda$. It is a factor of 30 more precise than a limit that has been derived~\cite{Baessler2007} from our previous experiment with ultra-cold neutrons~\cite{Nesvizhevsky2005,Westphal2007}. Other experiments have searched for spin-mass coupling at larger $\lambda$ and extrapolated over several orders of magnitude. A recent review~\cite{Dubbers2011a} is covering this topic. The most restrictive limit on the product $g_sg_p/\hbar c$ has been derived by combining the existing laboratory limit on the scalar interaction~$g_s$ with stellar energy-loss limits on the pseudoscalar coupling $g_p$, which is more restrictive than laboratory searches~\cite{Raffelt2012}.

Our method profits from small systematic effects in such systems, mainly owing to the fact that, in contrast to atoms, the electric polarizability of neutrons is extreme low. Our error is dominated by the statistical uncertainty. Our experiment leads the way in the application of GRS to probe for new particle physics and to search for Non-Newtonian gravity with high precision.
 
\section{Methods}
\begin{figure}[t]
	\centering
		\includegraphics[width=0.45\textwidth]{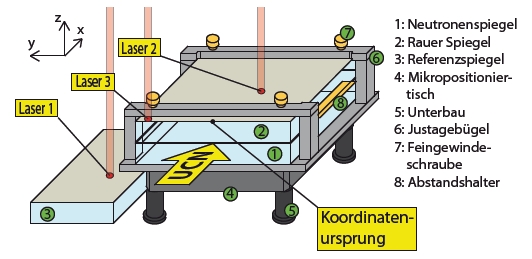}
		\includegraphics[width=0.45\textwidth]{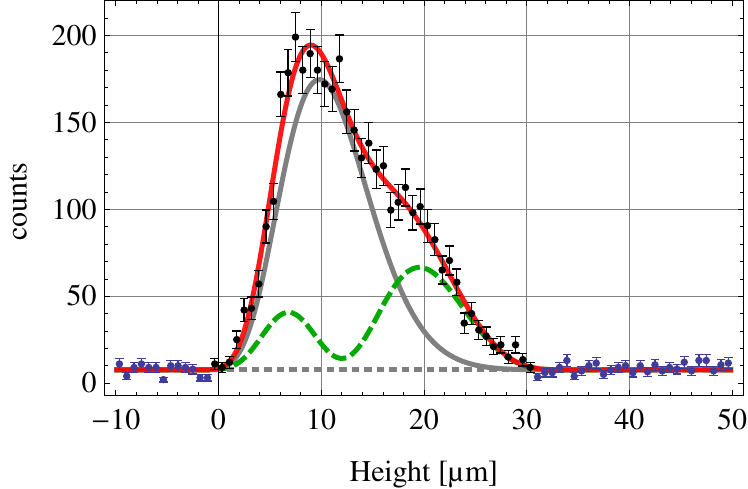}
	\caption{Left: a schematic diagram of the GRS setup. 1: bottom neutron mirror; 2: top rough mirror; 3: reference mirror; 4: micro-positioning table; 5: piezo-control; 6:adjustment; 7: fine pitch thread; 8: spacer. Right: neutron density distribution above the mirror in front of the detector, measured by a track detector~\cite{Jenke2009,Ruess2000,Nahrwold2004,Jenke2008}. A fit to the data (red) is compatible with a population of 67$\%$ for state $|1\rangle$ (grey) and 33$\%$ for state $|2\rangle$ (green).}
	\label{fig:lins}
\end{figure}

The setup is shown in Fig.~\ref{fig:lins}. Gravity Resonance Spectroscopy is a Rabi-type spectroscopy method and consists of a state selector, where the initial state $|p\rangle$ is prepared. By applying  a so-called  $\pi$-pulse, transitions into a second state $|q\rangle$ are induced. A detector at the end measures these neutrons, which were accepted by the state selector. On resonance a sharp decrease in count rate is found. The frequency width of the resonance is given by the time the neutron spends in the oscillating system. In our case the resonant transitions $|1\rangle \leftrightarrow |2\rangle$ and $|2\rangle \leftrightarrow |3\rangle$ are not separated; there exists a frequency band, where transitions $|1\rangle \leftrightarrow |2\rangle \leftrightarrow |3\rangle$ can simultaneously be driven, but they are in frequency space well separated from $|1\rangle \leftrightarrow |3\rangle$ transitions.  The setup is consisting of a polished mirror on the bottom and a rough scatterer on top. In addition a third mirror serves as reference. Vibrations are controlled with a three-beam laser interferometer.

\begin{figure}[tb]
	\centering
		\includegraphics[width=0.60\textwidth]{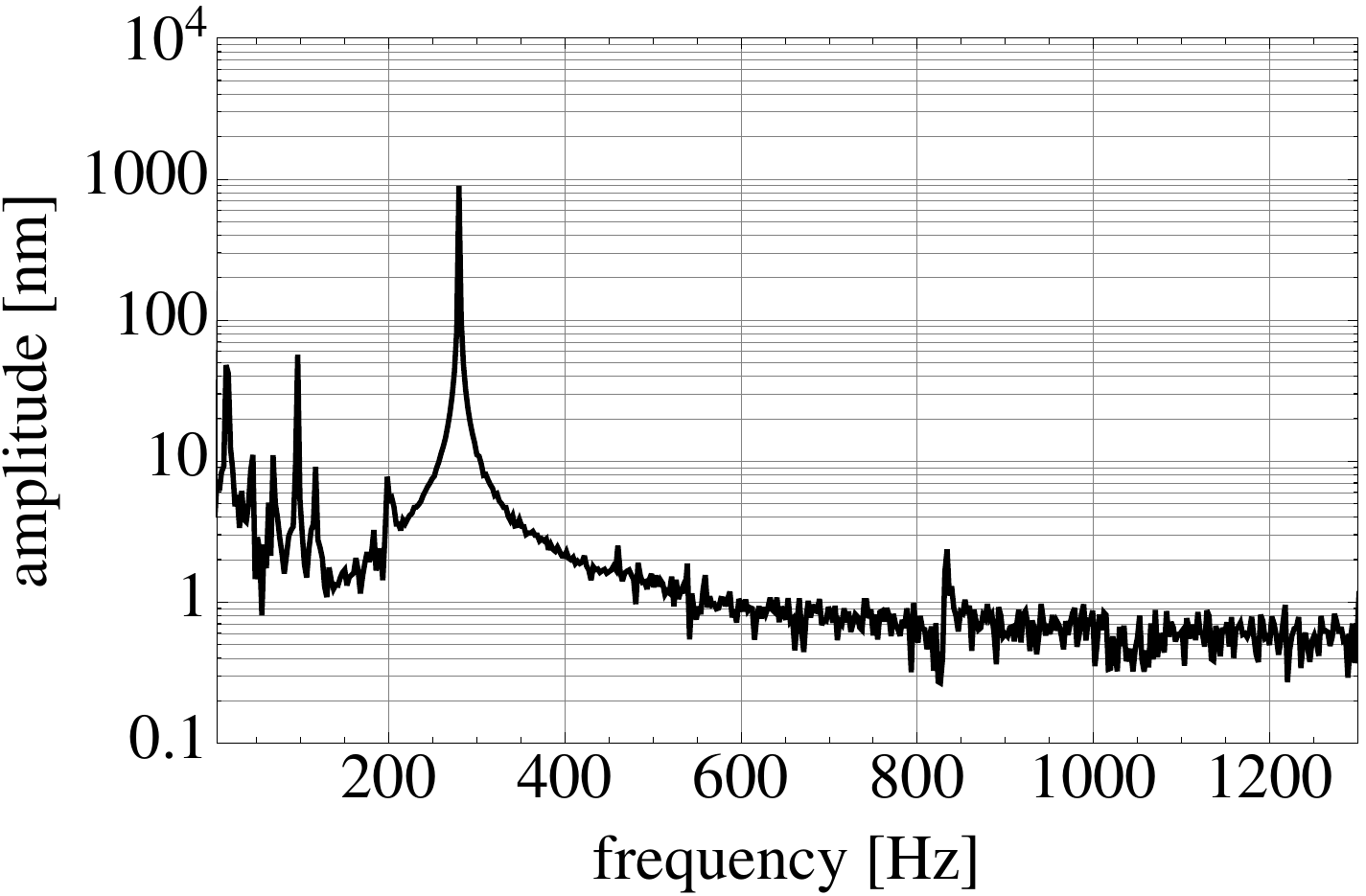}\\
		\includegraphics[width=0.45\textwidth]{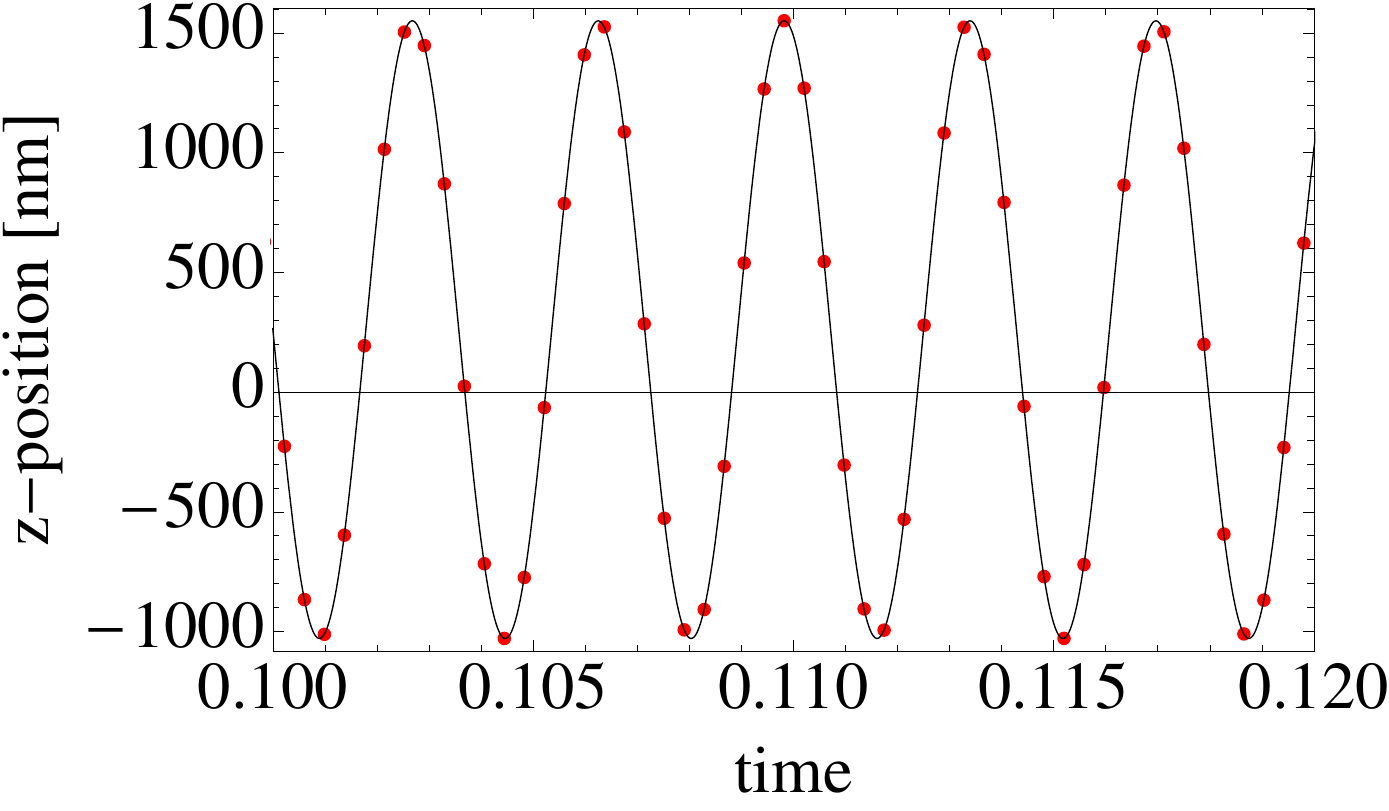}
		\includegraphics[width=0.45\textwidth]{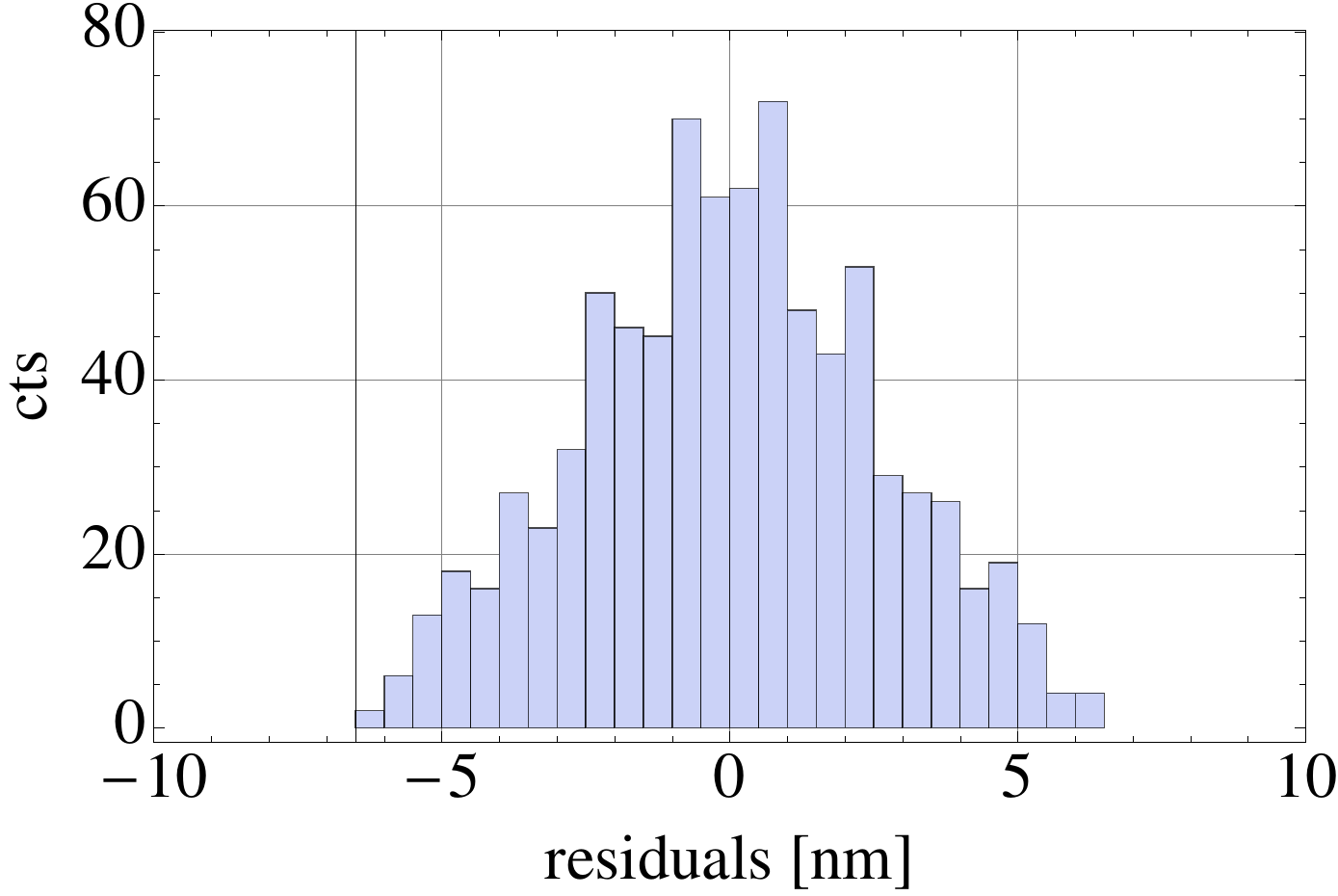}
	\caption{Top: FFT of a typical frequency measurement with a sampling rate of 2604~s$^{-1}$. The frequency  is 280~Hz at a vibration amplitude of 2.12~mm/s. Resonant frequencies of the mirror setup at 120~Hz and of the turbomolecular pump at 833~Hz are visible.
Left: Inverse FFT and a fit to vibration scan. Right: Residuals for amplitude measurement.}
	\label{fig:FFT}
\end{figure}
A neutron entering this system is in a superposition of different eigenstates $\varphi_n$. The wavefunction as a solution of the Schr\"{o}dinger equation between the mirrors with distance $l$, acceleration $g$ and height $z$ is given by
\beq
\psi\left( z,l,g\right) = \sum_n c_n e^{-\gamma_n/2\cdot t} e^{-i\phi_n} \varphi_n\left( z,l,g \right).
\eeq
$c_n$ is the state population of state $n$ at the entrance of the device, the damping term $e^{-\gamma_n/2t}$ takes the damping of state $n$ during the time $t$ of flight into account, where the factor $\gamma_n$ depends on the roughness of the upper mirror, and $e^{-i\phi_n}$ is a phase factor. When the vibrations are turned off, the population in front of the detector is 67$\%$ for state $|1\rangle$, and 33\% for state $|2\rangle$, state $|3\rangle$ and higher states are zero. The state selector is a mirror with a rough surface of roughness 3~{\textmu}m, and suppresses the population of higher levels by scattering and absorbing them. At the beginning of the experiment, the state population has been extracted from  $\left|\psi(z,l,g)\right|^2$, the spatial neutron distribution above the mirror i.e.~measured with a track detector~\cite{Ruess2000,Nahrwold2004,Jenke2008}, see Fig.~\ref{fig:lins}, right.   $\left|\psi(z,l,g)\right|^2$ is an incoherent superposition of state $|1\rangle$ and $|2\rangle$ since the phase is not defined. For the spectroscopy measurements described in this paper, this spatial resolution detector is replaced by a counting detector. In that detector, the entrance window is made out of aluminium coated with $^{10}$B, converting the neutron to $^7$Li and $\alpha$ with an efficiency of about 92\%. The background rate is $N_0 = \left( 2.18 \pm 0.08 \right)\times 10^{-3}$~s$^{-1}$. Background measurements are done every 200~s and in addition six longtime measurements are performed. The signal is registered with a usual preamplifier, main-amplifier and multi-channel analyzer chain. In addition, a SADC-analyzer measures the identical neutron signal, allowing a post procession of the data due to energy and time registration as a function of time. For spin dependent searches, the detector foil is replaced by a foil coated with soft iron on the other side serving as an analyser for the direction of the spin of the neutron. The degree of the foil polarization has been measured separately to be ca.~93$\%$. During all measurement, a beam monitor registered the neutron flux. The horizontal neutron velocity is $\left( 6 < v_x < 10 \right)$~m/s, which was adjusted with baffles in front of the setup.

For the Rabi spectroscopy measurement, the observable is the measured transmission as a function of the modulation frequency and amplitude. This is done using a vibrating mirror i.e. a modulation of the mirror's vertical position. For this purpose, piezo-elements are attached underneath, see Fig~\ref{fig:lins}. They induce a rapid modulation of the surface height with amplitude $A$.\\
In accordance with transmission measurements of Fig.~\ref{fig:GRS}, resonant transitions are possible between, $|1\rangle \leftrightarrow  |2\rangle$, $|2\rangle \leftrightarrow  |3\rangle$ ,$|2\rangle \leftrightarrow  |4\rangle$, and $|1\rangle \leftrightarrow  |3\rangle$. The resonance width of about 53~Hz causes simultaneous transitions between  $|1\rangle \leftrightarrow  |2\rangle$ and $ |2\rangle  \leftrightarrow |3\rangle$, they "overlap". As a consequence the transitions are treated as three-level systems and the time evolution of the states are given by the following differential equation:
\begin{align}
	\left(\begin{array}{c} \dot{b}_1(t) \\ \dot{b}_2(t) \\ \dot{b}_3(t) \\ \dot{b}_4(t) \end{array}\right)
	&=
	\left(\begin{array}{cccc}
		-\frac{\gamma_1}{2} & - S_{21}			& 0					& 0\\
			S_{21}		& -\frac{\gamma_2}{2} 	& - S_{32} 			& 0\\
			0			& S_{32}				&-\frac{\gamma_3}{2}	& 0\\
			0			& 0					& 0					& 0
	\end{array}\right) \cdot
	\left(\begin{array}{c} b_1(t) \\ b_2(t) \\ b_3(t) \\ b_3(t)\end{array}\right)\\
	\left(\begin{array}{c} \dot{b}_1(t) \\ \dot{b}_2(t) \\ \dot{b}_3(t) \\ \dot{b}_4(t) \end{array}\right)
	&=
	\left(\begin{array}{cccc}
		-\frac{\gamma_1}{2} & 0					& - S_{31}			& 0\\
			0			& -\frac{\gamma_2}{2} 	& 0		 			& - S_{42}\\
			S_{31}		& 0					&-\frac{\gamma_3}{2}	& 0\\
			0			& S_{42}				& 0					& -\frac{\gamma_4}{2}
	\end{array}\right) \cdot
	\left(\begin{array}{c} b_1(t) \\ b_2(t) \\ b_3(t) \\ b_3(t)
\end{array}\right)\\
	\label{eq:grs-rabi-34}
	\end{align}
Here, for two states $|p\rangle$ and $|q\rangle$ with relative phase~$\phi_{qp}$, $S_{qp}$ is given by
\beq
S_{qp} = \frac{1}{2} e^{-i \phi_{qp}} e^{-i \delta_{qp} t} d \langle q | \frac{\partial}{\partial z} | p \rangle.
\eeq
The frequency detuning from resonance reads $\delta_{qp} = 2\pi \left( \nu - \nu_{qp} \right)$. The vibration amplitude
in units of a velocity is given by~$u = 2\pi\nu a_k$.

The first equation examines transitions $|1\rangle  \leftrightarrow |2\rangle$ and $|2\rangle  \leftrightarrow |3\rangle$ for states $|1\rangle$, $|2\rangle$, $|3\rangle$, whereas the second equation system examines $|1\rangle \leftrightarrow  |3\rangle$ and $|2\rangle \leftrightarrow  |4\rangle$ for states $|1\rangle$, $|2\rangle$, $|3\rangle$, and $|4\rangle$, which is simpler because these transitions do not overlap.

Vibration control is done in a twofold way.  Firstly, a four-channel acceleration sensor-system measures external vibrations at different positions for surveillance. Secondly, a three-beam laser interferometer is in use. Laser beam~$L1$ serves as a reference beam, which measures vibrations on a separate mirror placed directly on the granite table. Laser~$L2$ scans the surface of the vibrating neutron mirror to yield a spatial- and time-dependent map. 
The total deviation of the vibration amplitude from its mean was usually less than 20\%.
$L3$ scans for vibrations at a fixed position on the neutron mirror.
From this data, a vibration analysis to extract the vibration amplitude~$a_k$ and relative phase at different positions of the mirror surface is performed using the expression
\beq
f\left( N_0, a_k, \nu_k, \phi_k\right) = N_0 + \sum_{k=1}^{K} a_k \cdot \sin{2 \pi \nu_k \cdot t+\phi_k},
\label{eq:fitvib01}
\eeq
where $N_0$ is a constant offset,  amplitude $a_k$, for frequency  $\nu_k$ and Phase $\phi_k$. The number of fit parameters are $3K + 1$, where $k$ is the number of sine-modes.

 In all measurements, a linear gradient of the vibration amplitude on the surface was found, which has no influence on the Gravity Resonance Spectroscopy method to first order.

\begin{figure}[t]
	\centering
		\includegraphics[width=0.45\textwidth]{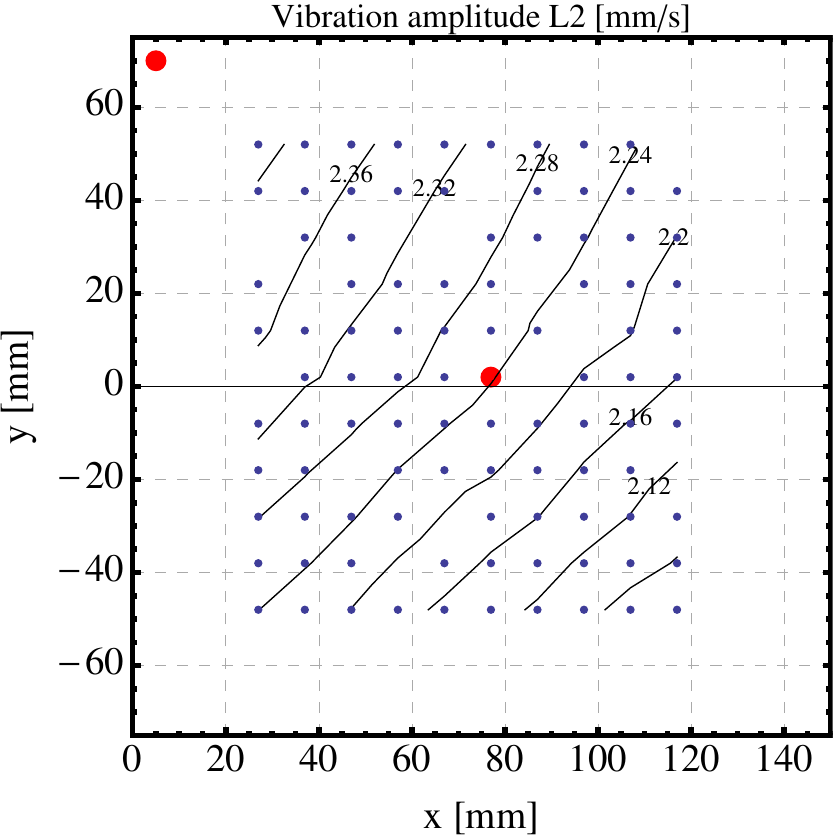}
		\includegraphics[width=0.45\textwidth]{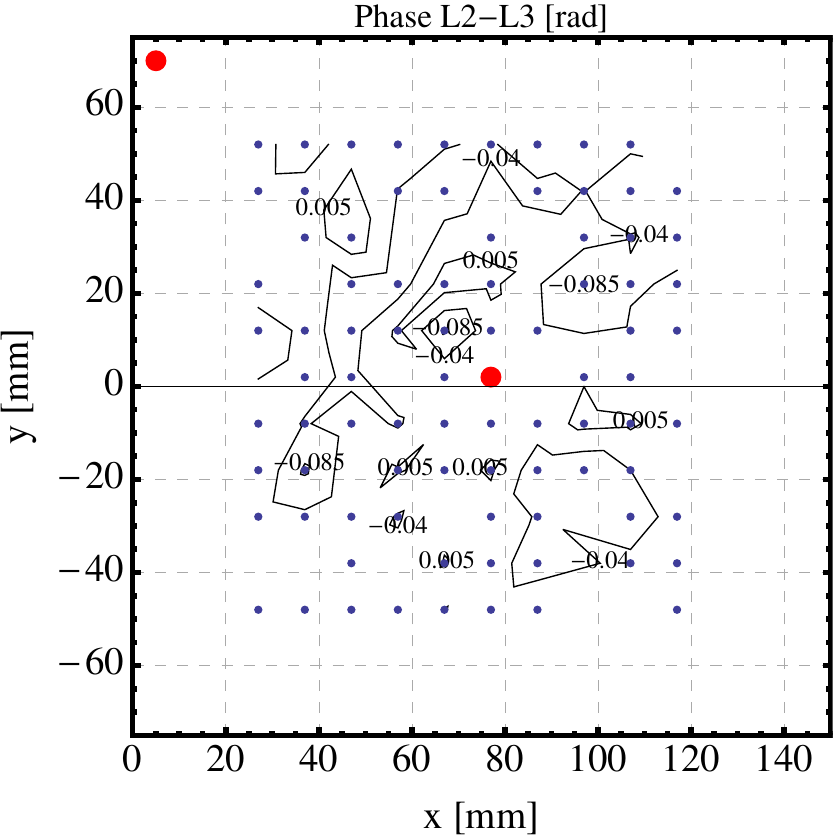}
	\caption{Two-dimensional map of the vibrating mirror for amplitude (left) and phase (right).\newline
A grid of 12~$\times$~10 (blue) data points with each 2$^{13}$ data samples taken at a sampling rate of 2604~s$^{-1}$ results in a frequency resolution of 10$^{-4}$. The left figure shows, that the vibration amplitude is homogeneous on the 10\%-level.
Its relative phase is close to zero, as the plot on the right indicates.
The red labels indicate the laser beams $L2$ and $L3$.}
	\label{fig:vibphase}
\end{figure}

A typical frequency measurement is shown in Fig~\ref{fig:FFT}, top. After fast fourier transformation (FFT) a gauss filter is applied and inverse FFT in order to keep computing time small for adjustment to the sine-function, see Fig.~\ref{fig:FFT}, left and right.
A two-dimensional map of the vibrating mirror for amplitude and frequency is shown in Fig.~\ref{fig:vibphase}. 

\vspace{2cm}
\begin{acknowledgments}
We thank Larisa Chizhova, Hanno Filter, Guillaume Pignol and Stefan Rotter for useful discussions, Dominik Seiler, TU Munich, for the preparation of the neutron spatial resolution detectors and Rainer Stadler, University of Heidelberg, for the preparation of the rough mirror surface. The mirrors by itself were characterized by S--DH Sputterd{\"u}nnschichttechnik, Heidelberg. We thank Thomas Brenner, ILL, and Ralf Ziegler, University of Heidelberg, for technical support. We gratefully acknowledge support from the Austrian \textit{Fonds zur F\"orderung der Wissenschaftlichen Forschung} (FWF) under Contract No. I529-N20 and the German Research Foundation (DFG) as part of  the Priority Programme (SPP) 1491 "Precision experiments in particle and astrophysics with cold and ultracold neutrons"; we also greatfully acknowledge support from the French \textit{L'Agence nationale de la recherche} ANR under contract number ANR-2011-ISO4-007-02, Programme Blanc International - SIMI4-Physique.
\end{acknowledgments}

\bibliography{Referenzen}{}

\begin{thebibliography}{10}
\expandafter\ifx\csname url\endcsname\relax
  \def\url#1{\texttt{#1}}\fi
\expandafter\ifx\csname urlprefix\endcsname\relax\def\urlprefix{URL }\fi
\providecommand{\bibinfo}[2]{#2}
\providecommand{\eprint}[2][]{\url{#2}}

\bibitem{LHC2012}
\bibinfo{author}{Gillies, J.}
\newblock \bibinfo{title}{Cern press release, 4 july 2012}
  (\bibinfo{year}{2012}).

\bibitem{atlas}
\bibinfo{author}{{The ATLAS Collaboration}}.
\newblock \bibinfo{title}{{Observation of a new particle in the search for the
  Standard Model Higgs boson with the ATLAS detector at the LHC}}
  \bibinfo{pages}{24} (\bibinfo{year}{2012}).
\newblock \eprint{1207.7214}.

\bibitem{cms}
\bibinfo{author}{{The CMS Collaboration}}.
\newblock \bibinfo{title}{{Observation of a new boson at a mass of 125 GeV with
  the CMS experiment at the LHC}}  (\bibinfo{year}{2012}).
\newblock \eprint{1207.7235}.

\bibitem{Adelberger2009}
\bibinfo{author}{Adelberger, E.}, \bibinfo{author}{Gundlach, J.},
  \bibinfo{author}{Heckel, B.}, \bibinfo{author}{Hoedl, S.} \&
  \bibinfo{author}{Schlamminger, S.}
\newblock \bibinfo{title}{{Torsion balance experiments: A low-energy frontier
  of particle physics}}.
\newblock \emph{\bibinfo{journal}{Progress in Particle and Nuclear Physics}}
  \textbf{\bibinfo{volume}{62}}, \bibinfo{pages}{102--134}
  (\bibinfo{year}{2009}).

\bibitem{Brax2011}
\bibinfo{author}{Brax, P.} \& \bibinfo{author}{Pignol, G.}
\newblock \bibinfo{title}{{Strongly Coupled Chameleons and the Neutronic
  Quantum Bouncer}}.
\newblock \emph{\bibinfo{journal}{Physical Review Letters}}
  \textbf{\bibinfo{volume}{107}}, \bibinfo{pages}{1--4} (\bibinfo{year}{2011}).

\bibitem{Jenke2011}
\bibinfo{author}{Jenke, T.}, \bibinfo{author}{Geltenbort, P.},
  \bibinfo{author}{Lemmel, H.} \& \bibinfo{author}{Abele, H.}
\newblock \bibinfo{title}{{Realization of a gravity-resonance-spectroscopy
  technique}}.
\newblock \emph{\bibinfo{journal}{Nature Physics}}
  \textbf{\bibinfo{volume}{7}}, \bibinfo{pages}{468--472}
  (\bibinfo{year}{2011}).

\bibitem{Nesvizhevsky2005}
\bibinfo{author}{Nesvizhevsky, V.~V.} \emph{et~al.}
\newblock \bibinfo{title}{{Study of the neutron quantum states in the gravity
  field}}.
\newblock \emph{\bibinfo{journal}{Eur. Phys. J.}}
  \textbf{\bibinfo{volume}{C40}}, \bibinfo{pages}{479--491}
  (\bibinfo{year}{2005}).

\bibitem{Westphal2007}
\bibinfo{author}{Westphal, A.} \emph{et~al.}
\newblock \bibinfo{title}{{A quantum mechanical description of the experiment
  on the observation of gravitationally bound states}}.
\newblock \emph{\bibinfo{journal}{The European Physical Journal C}}
  \textbf{\bibinfo{volume}{51}}, \bibinfo{pages}{367--375}
  (\bibinfo{year}{2007}).

\bibitem{Riess1998}
\bibinfo{author}{Riess, A.} \emph{et~al.}
\newblock \bibinfo{title}{Observational evidence from supernovae for an
  accelerating universeand a cosmological constant}.
\newblock \emph{\bibinfo{journal}{The Astronomical Journal}}
  \textbf{\bibinfo{volume}{116}} (\bibinfo{year}{1998}).

\bibitem{Lahav2010}
\bibinfo{author}{Lahav, O.} \& \bibinfo{author}{Liddle, A.~R.}
\newblock \bibinfo{title}{{The Cosmological Parameters 2010}}
  \bibinfo{pages}{25} (\bibinfo{year}{2010}).
\newblock \eprint{1002.3488}.

\bibitem{Glashow1970}
\bibinfo{author}{Glashow, S.~L.}, \bibinfo{author}{Iliopoulos, J.} \&
  \bibinfo{author}{Maiani, L.}
\newblock \bibinfo{title}{{Weak Interactions with Lepton-Hadron Symmetry}}.
\newblock \emph{\bibinfo{journal}{Physical Review D}}
  \textbf{\bibinfo{volume}{2}}, \bibinfo{pages}{1285--1292}
  (\bibinfo{year}{1970}).

\bibitem{Weinberg1967}
\bibinfo{author}{Weinberg, S.}
\newblock \bibinfo{title}{{A Model of Leptons}}.
\newblock \emph{\bibinfo{journal}{Physical Review Letters}}
  \textbf{\bibinfo{volume}{19}}, \bibinfo{pages}{1264--1266}
  (\bibinfo{year}{1967}).

\bibitem{Salam1969}
\bibinfo{author}{Salam, A.}
\newblock \emph{\bibinfo{title}{{Elementary Particle Theory}}}
  (\bibinfo{publisher}{Almquist and Wiksells, Stockholm},
  \bibinfo{year}{1969}).

\bibitem{Silvestri2009}
\bibinfo{author}{Silvestri, A.} \& \bibinfo{author}{Trodden, M.}
\newblock \bibinfo{title}{{Approaches to understanding cosmic acceleration}}.
\newblock \emph{\bibinfo{journal}{Reports on Progress in Physics}}
  \textbf{\bibinfo{volume}{72}}, \bibinfo{pages}{096901}
  (\bibinfo{year}{2009}).

\bibitem{Caldwell2009}
\bibinfo{author}{Caldwell, R.~R.} \& \bibinfo{author}{Kamionkowski, M.}
\newblock \bibinfo{title}{{The Physics of Cosmic Acceleration}}.
\newblock \emph{\bibinfo{journal}{Annual Review of Nuclear and Particle
  Science}} \textbf{\bibinfo{volume}{59}}, \bibinfo{pages}{397--429}
  (\bibinfo{year}{2009}).

\bibitem{Copeland2006}
\bibinfo{author}{Copeland, E.~J.}, \bibinfo{author}{Sami, M.} \&
  \bibinfo{author}{Tsujikawa, S.}
\newblock \bibinfo{title}{Dynamics of dark energy}.
\newblock \emph{\bibinfo{journal}{International Journal of Modern Physics D}}
  \textbf{\bibinfo{volume}{15}}, \bibinfo{pages}{1753--1935}
  (\bibinfo{year}{2006}).

\bibitem{Wetterich1988}
\bibinfo{author}{Wetterich, C.}
\newblock \bibinfo{title}{{Cosmology and the fate of dilatation symmetry}}.
\newblock \emph{\bibinfo{journal}{Nuclear Physics B}}
  \textbf{\bibinfo{volume}{302}}, \bibinfo{pages}{668--696}
  (\bibinfo{year}{1988}).

\bibitem{Beringer2012}
\bibinfo{author}{Beringer, J.} \emph{et~al.}
\newblock \bibinfo{title}{{Review of Particle Physics}}.
\newblock \emph{\bibinfo{journal}{Physical Review D}}
  \textbf{\bibinfo{volume}{86}} (\bibinfo{year}{2012}).

\bibitem{Arkani-Hamed1999}
\bibinfo{author}{Arkani-Hamed, N.}, \bibinfo{author}{Dimopoulos, S.} \&
  \bibinfo{author}{Dvali, G.}
\newblock \bibinfo{title}{{Phenomenology, astrophysics, and cosmology of
  theories with submillimeter dimensions and TeV scale quantum gravity}}.
\newblock \emph{\bibinfo{journal}{Physical Review D}}
  \textbf{\bibinfo{volume}{59}}, \bibinfo{pages}{1--21} (\bibinfo{year}{1999}).

\bibitem{Abele2003}
\bibinfo{author}{Abele, H.}, \bibinfo{author}{Bae{\ss}ler, S.} \&
  \bibinfo{author}{Westphal, A.}
\newblock \bibinfo{title}{{Quantum states of neutrons in the gravitational
  field and limits for non-Newtonian interaction in the range between 1 micron
  and 10 microns}} \bibinfo{pages}{1--13} (\bibinfo{year}{2003}).
\newblock \eprint{0301145}.

\bibitem{Fischbach1999}
\bibinfo{author}{E~Fischbach, C. L.~T.}
\newblock \emph{\bibinfo{title}{{The Search for Non-Newtonian Gravity}}}
  (\bibinfo{publisher}{Springer-Verlag, New York}, \bibinfo{year}{1999}).

\bibitem{Tu2007}
\bibinfo{author}{Tu, L.-C.}, \bibinfo{author}{Guan, S.-G.},
  \bibinfo{author}{Luo, J.}, \bibinfo{author}{Shao, C.-G.} \&
  \bibinfo{author}{Liu, L.-X.}
\newblock \bibinfo{title}{{Null Test of Newtonian Inverse-Square Law at
  Submillimeter Range with a Dual-Modulation Torsion Pendulum}}.
\newblock \emph{\bibinfo{journal}{Physical Review Letters}}
  \textbf{\bibinfo{volume}{98}} (\bibinfo{year}{2007}).

\bibitem{Hoyle2004}
\bibinfo{author}{Hoyle, C.~D.} \emph{et~al.}
\newblock \bibinfo{title}{{Sub-millimeter Tests of the Gravitational
  Inverse-square Law}}.
\newblock \emph{\bibinfo{journal}{Physical Review D}}
  \textbf{\bibinfo{volume}{70}}, \bibinfo{pages}{34} (\bibinfo{year}{2004}).
\newblock \eprint{0405262}.

\bibitem{Decca2005}
\bibinfo{author}{Decca, R.} \emph{et~al.}
\newblock \bibinfo{title}{{Constraining New Forces in the Casimir Regime Using
  the Isoelectronic Technique}}.
\newblock \emph{\bibinfo{journal}{Physical Review Letters}}
  \textbf{\bibinfo{volume}{94}} (\bibinfo{year}{2005}).

\bibitem{Long2003}
\bibinfo{author}{Long, J.~C.} \emph{et~al.}
\newblock \bibinfo{title}{{Upper limits to submillimetre-range forces from
  extra space-time dimensions.}}
\newblock \emph{\bibinfo{journal}{Nature}} \textbf{\bibinfo{volume}{421}},
  \bibinfo{pages}{922--5} (\bibinfo{year}{2003}).

\bibitem{Geraci2008}
\bibinfo{author}{Geraci, A.}, \bibinfo{author}{Smullin, S.},
  \bibinfo{author}{Weld, D.}, \bibinfo{author}{Chiaverini, J.} \&
  \bibinfo{author}{Kapitulnik, A.}
\newblock \bibinfo{title}{{Improved constraints on non-Newtonian forces at 10
  microns}}.
\newblock \emph{\bibinfo{journal}{Physical Review D}}
  \textbf{\bibinfo{volume}{78}}, \bibinfo{pages}{12} (\bibinfo{year}{2008}).
\newblock \eprint{0802.2350}.

\bibitem{Chiaverini2003}
\bibinfo{author}{Chiaverini, J.}, \bibinfo{author}{Smullin, S.},
  \bibinfo{author}{Geraci, A.}, \bibinfo{author}{Weld, D.} \&
  \bibinfo{author}{Kapitulnik, A.}
\newblock \bibinfo{title}{{New Experimental Constraints on Non-Newtonian Forces
  below 100 $\mu$m}}.
\newblock \emph{\bibinfo{journal}{Physical Review Letters}}
  \textbf{\bibinfo{volume}{90}} (\bibinfo{year}{2003}).

\bibitem{Bordag2001}
\bibinfo{author}{Bordag, M.}, \bibinfo{author}{Mohideen, U.} \&
  \bibinfo{author}{Mostepanenko, V.}
\newblock \bibinfo{title}{{New developments in the Casimir effect}}.
\newblock \emph{\bibinfo{journal}{Physics Reports}}
  \textbf{\bibinfo{volume}{353}}, \bibinfo{pages}{1--205}
  (\bibinfo{year}{2001}).

\bibitem{Leeb1992}
\bibinfo{author}{Leeb, H.} \& \bibinfo{author}{Schmiedmayer, J.}
\newblock \bibinfo{title}{{Constraint on hypothetical light interacting bosons
  from low-energy neutron experiments.pdf}}.
\newblock \emph{\bibinfo{journal}{Physical Review Letters}}
  \textbf{\bibinfo{volume}{68}} (\bibinfo{year}{1992}).

\bibitem{Nesvizhevsky2008}
\bibinfo{author}{Nesvizhevsky, V.}, \bibinfo{author}{Pignol, G.} \&
  \bibinfo{author}{Protasov, K.}
\newblock \bibinfo{title}{{Neutron scattering and extra-short-range
  interactions}}.
\newblock \emph{\bibinfo{journal}{Physical Review D}}
  \textbf{\bibinfo{volume}{77}} (\bibinfo{year}{2008}).

\bibitem{Dubbers2011a}
\bibinfo{author}{Dubbers, D.} \& \bibinfo{author}{Schmidt, M.}
\newblock \bibinfo{title}{{The neutron and its role in cosmology and particle
  physics}}.
\newblock \emph{\bibinfo{journal}{Reviews of Modern Physics}}
  \textbf{\bibinfo{volume}{83}}, \bibinfo{pages}{1111--1171}
  (\bibinfo{year}{2011}).

\bibitem{Jenke2011a}
\bibinfo{author}{Jenke, T.}
\newblock \bibinfo{title}{Dissertation, {T}echnische {U}niverst\"at {W}ien}
  (\bibinfo{year}{2011}).

\bibitem{Kreuz2009}
\bibinfo{author}{Kreuz, M.} \emph{et~al.}
\newblock \bibinfo{title}{{A method to measure the resonance transitions
  between the gravitationally bound quantum states of neutrons in the GRANIT
  spectrometer}}  (\bibinfo{year}{2009}).
\newblock \eprint{0902.0156}.

\bibitem{Khoury2004}
\bibinfo{author}{Khoury, J.} \& \bibinfo{author}{Weltman, A.}
\newblock \bibinfo{title}{{Chameleon Fields: Awaiting Surprises for Tests of
  Gravity in Space}}.
\newblock \emph{\bibinfo{journal}{Physical Review Letters}}
  \textbf{\bibinfo{volume}{93}} (\bibinfo{year}{2004}).

\bibitem{Mota2006}
\bibinfo{author}{Mota, D.~F.} \& \bibinfo{author}{Shaw, D.~J.}
\newblock \bibinfo{title}{{Strongly Coupled Chameleon Fields: New Horizons in
  Scalar Field Theory}}.
\newblock \emph{\bibinfo{journal}{Physical Review Letters}}
  \textbf{\bibinfo{volume}{97}} (\bibinfo{year}{2006}).

\bibitem{Mota2007}
\bibinfo{author}{Mota, D.} \& \bibinfo{author}{Shaw, D.}
\newblock \bibinfo{title}{{Evading equivalence principle violations,
  cosmological, and other experimental constraints in scalar field theories
  with a strong coupling to matter}}.
\newblock \emph{\bibinfo{journal}{Physical Review D}}
  \textbf{\bibinfo{volume}{75}} (\bibinfo{year}{2007}).

\bibitem{Waterhouse2006}
\bibinfo{author}{Waterhouse, T.~P.}
\newblock \bibinfo{title}{{An Introduction to Chameleon Gravity}}
  \bibinfo{pages}{33} (\bibinfo{year}{2006}).
\newblock \eprint{0611816}.

\bibitem{Ahlers2008}
\bibinfo{author}{Ahlers, M.}, \bibinfo{author}{Lindner, A.},
  \bibinfo{author}{Ringwald, A.}, \bibinfo{author}{Schrempp, L.} \&
  \bibinfo{author}{Weniger, C.}
\newblock \bibinfo{title}{{Alpenglow: A signature for chameleons in axionlike
  particle search experiments}}.
\newblock \emph{\bibinfo{journal}{Physical Review D}}
  \textbf{\bibinfo{volume}{77}} (\bibinfo{year}{2008}).

\bibitem{Gies2008}
\bibinfo{author}{Gies, H.}, \bibinfo{author}{Mota, D.} \&
  \bibinfo{author}{Shaw, D.}
\newblock \bibinfo{title}{{Hidden in the light: Magnetically induced afterglow
  from trapped chameleon fields}}.
\newblock \emph{\bibinfo{journal}{Physical Review D}}
  \textbf{\bibinfo{volume}{77}} (\bibinfo{year}{2008}).

\bibitem{Steffen2010}
\bibinfo{author}{Steffen, J.} \emph{et~al.}
\newblock \bibinfo{title}{{Laboratory Constraints on Chameleon Dark Energy and
  Power-Law Fields}}.
\newblock \emph{\bibinfo{journal}{Physical Review Letters}}
  \textbf{\bibinfo{volume}{105}} (\bibinfo{year}{2010}).

\bibitem{Upadhye2012}
\bibinfo{author}{Upadhye, A.}, \bibinfo{author}{Steffen, J.} \&
  \bibinfo{author}{Chou, A.}
\newblock \bibinfo{title}{{Designing dark energy afterglow experiments}}.
\newblock \emph{\bibinfo{journal}{Physical Review D}}
  \textbf{\bibinfo{volume}{86}}, \bibinfo{pages}{1--29} (\bibinfo{year}{2012}).

\bibitem{Rybka2010}
\bibinfo{author}{Rybka, G.} \emph{et~al.}
\newblock \bibinfo{title}{{Search for Chameleon Scalar Fields with the Axion
  Dark Matter Experiment}}.
\newblock \emph{\bibinfo{journal}{Physical Review Letters}}
  \textbf{\bibinfo{volume}{105}} (\bibinfo{year}{2010}).

\bibitem{Wester2011}
\bibinfo{author}{Wester, W.}
\newblock \bibinfo{title}{Laser experiments at fermilab for wisps and other
  effects} (\bibinfo{publisher}{Presentation at the 7th Patras Workshop on
  Axions, WIMPs and WISPs, 2011.}, \bibinfo{year}{2011}).

\bibitem{Ivanov2012}
\bibinfo{author}{Ivanov, A.~N.}, \bibinfo{author}{H\"{o}llwieser, R.},
  \bibinfo{author}{Jenke, T.}, \bibinfo{author}{Wellenzohen, M.} \&
  \bibinfo{author}{Abele, H.}
\newblock \bibinfo{title}{{The influence of the chameleon field potential on
  transition frequencies of gravitationally bound quantum states of ultra-cold
  neutrons}} \bibinfo{pages}{11} (\bibinfo{year}{2012}).
\newblock \eprint{1207.0419}.

\bibitem{Moody1984}
\bibinfo{author}{Moody, J.} \& \bibinfo{author}{Wilczek, F.}
\newblock \bibinfo{title}{{New macroscopic forces?}}
\newblock \emph{\bibinfo{journal}{Physical Review D}}
  \textbf{\bibinfo{volume}{30}}, \bibinfo{pages}{130--138}
  (\bibinfo{year}{1984}).

\bibitem{Baessler2007}
\bibinfo{author}{Bae{\ss}ler, S.}, \bibinfo{author}{Nesvizhevsky, V.},
  \bibinfo{author}{Protasov, K.} \& \bibinfo{author}{Voronin, A.}
\newblock \bibinfo{title}{{Constraint on the coupling of axionlike particles to
  matter via an ultracold neutron gravitational experiment}}.
\newblock \emph{\bibinfo{journal}{Physical Review D}}
  \textbf{\bibinfo{volume}{75}}, \bibinfo{pages}{1--4} (\bibinfo{year}{2007}).

\bibitem{Raffelt2012}
\bibinfo{author}{Raffelt, G.}
\newblock \bibinfo{title}{Limits on a {CP}-violating scalar axion-nucleon
  interaction}.
\newblock \emph{\bibinfo{journal}{Physical Review D}}
  \textbf{\bibinfo{volume}{86}} (\bibinfo{year}{2012}).

\bibitem{Jenke2009}
\bibinfo{author}{Jenke, T.}, \bibinfo{author}{Stadler, D.},
  \bibinfo{author}{Abele, H.} \& \bibinfo{author}{Geltenbort, P.}
\newblock \bibinfo{title}{{Q-BOUNCE: Experiments with quantum bouncing
  ultracold neutrons}}.
\newblock \emph{\bibinfo{journal}{Nuclear Instruments and Methods in Physics
  Research Section A: Accelerators, Spectrometers, Detectors and Associated
  Equipment}} \textbf{\bibinfo{volume}{611}}, \bibinfo{pages}{318--321}
  (\bibinfo{year}{2009}).

\bibitem{Ruess2000}
\bibinfo{author}{Ruess, F.}
\newblock \bibinfo{title}{Diplomarbeit, {U}niverst\"at {H}eidelberg}
  (\bibinfo{year}{2000}).

\bibitem{Nahrwold2004}
\bibinfo{author}{Nahrwold, S.}
\newblock \bibinfo{title}{Diplomarbeit, {U}niverst\"at {H}eidelberg}
  (\bibinfo{year}{2004}).

\bibitem{Jenke2008}
\bibinfo{author}{Jenke, T.}
\newblock \bibinfo{title}{Diplomarbeit, {U}niverst\"at {H}eidelberg}
  (\bibinfo{year}{2008}).

\end{thebibliography}
\bibliographystyle{naturemag}

\end{document}